# The Economic Analysis of the Common Pool Method through the HARA Utility Functions

**Mu Lin [1], Di Zhang [1], Ben Chen [2], Hang Zheng [2,*]**

[1] Institute of statistics and mathematics, Central University of Finance and Economics, Changping district, China; 13260288776@163.com

[2] Hydraulic Engineering, Tsinghua University, Tsinghua University, Beijing, China ; zhenghang00@163.com

* Correspondence: zhenghang00@163.com

Water market is a contemporary marketplace for water trading and is deemed to one of the most efficient instruments to improve the social welfare. In modern water markets, the two widely used trading systems are an improved pair-wise trading, and a "smart market" or common pool method. In comparison with the economic model, this paper constructs a conceptual mathematic model through the HARA utility functions. Mirroring the concepts such as Nash Equilibrium, Pareto optimal and stable matching in economy, three significant propositions are acquired which illustrate the advantages of the common pool method compared with the improved pair-wise trading.

## 1. Introduction

Over the last decades, the increasing water demand against uneven water resources has been a significant problem [1]. Water market is a marketplace for water trading, which is regarded as one of the best instruments to provide gains in economic efficiency for water use [2,3]. By reallocating water from low-value use to higher-value use, the water market has been proved significantly in raising the social welfare in many countries and regions: on USA [4], on Australia; on Chile [5], on Aral Sea Basin [6]. Jordan A. Clayton [7] argues that regulated water markets provide flexible and just solutions to western water dilemmas, and reallocations may provide much-needed additional water supply. Water market system design is the key to the efficiency of water trading, an effective system which can optimize the welfare of the market, and provide much more gains for the traders.

There are two main water trading systems in the world, one is an improved pair-wise trading, and the other is a "smart market" or common pool method [8]. The pair-wise trading is current trading system in England. Under the pair-wise trading system, an abstractor wanting to buy or sell water rights must search for and make arrangements with another trader, the trade must be approved by the Environment Agency [8]. In Australia, the common pool method is the main trading instrument. In this method, the users buy and sell water rights with a catchment manger through a "common pool". By means of computerized market-clearing, the "common pool" can reduce transaction costs.

The "common pool" is based on smart market techniques, which uses computer models to assist market operation [9]. Murphy et al [10] proposed a smart market for water trading according to well-defined auction rules, where each user trades with auction manager [11]. The research on common pool has provided a series of insights for successful management of resources [12]. Since 1980s, the common pool of water trading has been increasingly used by farmers, environmental groups, governments, tourism providers, urban water providers, and industry and investment groups in Murray-Darling Basin, Australia [13]. A.M. zaman et al [14] use the water trading-allocation model to estimate the impact of temporary water trading and identify the economic benefits from temporary water trading. Alec Zuo et al [15] analyze the farm survey data from Australia's southern Murray-Darling basin, and find that purchasing water allocations can reduce the farm's risk on the market. Etchells T et al [16] give five separate strategies to overcome the third-part effect, facilitate market expansion and underpin investment confidence. In Australia, the water trading will halt ongoing decline in environmental conditions and resource security and provide a robust foundation for managing climate change [17]. Water market buyer clustering behavior was mostly explained by increased market uncertainty (in particular, hotter and drier conditions), while seller-clustering behavior is mostly explained by strategic behavioral factors which evaluate the costs and benefits of clustering [15].

The pair-wise trading system is a traditional method, where the traders search for each other and make arrangement by their own. It is the widely used water trading method in the world. In a pair wise trading system, the traders fall into different categories, where they could be either pre–approved or approved very quickly. It is an effective method to find appropriate trading partner. Erfani T et al [18] simulate water markets with transaction costs under the pair-wise system, the result show that buyers favor sellers who can sell larger volumes to minimize the number of transactions. So the trading cost is an important constraint of wise-pair trading method. Reddy S M W et al [19] establish the hydro-economic model to assess the water trading in Brazos river basin, finding that markets are unlikely to be a robust solution for Brazos because of the low reliability.

China is experiencing a combination of population growth, demographic change, and rapid economic development in recent years, which have dramatically increased demand for urban and industrial water use [20]. As an effective method of water management, water trading is necessary to respond to water scarcity of China. The comparison of benefits and shortages between trading and non-trading schemes implies that trading is more optimal and effective than non-trading [21]. The first attempt of water trading in China is the inter-jurisdictional transfers between two cities in Zhejiang Province, Yiwu and Dongyang, in 2000, which is viewed as one of the most successful market-based water management policies in China. Then, the inter-sectorial transfers between Ningxia and Inner Mongolia autonomous regions created the transfer from agricultural to industrial users [22]. Shuai Zhong et al [23] incorporated the water parallel pricing system of China within a computable general equilibrium (CGE) model, where the drought of 2000 is simulated and the results demonstrated that the effects on the macro-economy were insignificant. However, the effects on agricultural production, particularly on farming production mainly cultivated in northern areas were significant. Although Chinese water market has some processes, the common pool or the pair-wise, which type of water market system is appropriate for China. It remains a question.

Lumbroso D M et al [8] compared the common pool and pair-wise method by investigating traders in England. In their workshops, the traders are all English who get used of pair-wise method. In addition, their sample is too small to provide strong evidence. Robert Brooks and Edwyna Harris [24] find the common pool generates substantial economic benefits by empirical analysis of Water move data in Australia. Miguel Suarez Bosa [25] acquire the conclusion that user association management is successful (common pool method), whereas, individual management (pair-wise trading) can lead to squandering. Without enough data and investigation, which method should be taken for water trading?

The aforementioned researches concentrate on the empirical study, nonetheless, lack of analytic solution in theory[26], and the main cause is the disparate trading incentives among different market participants are difficult to measure. Although there exist scads of differences, we deliver the viewpoint that HARA (Hyperbolic Absolute Risk Aversion) utility function, which includes power utility, exponential utility, and logarithm utility as special cases [27], accords with common sense as well as our daily experience (continuity, monotone increasing and diminishing marginal utility), hence, is partially indisputable. The more concrete details pertain HARA are indicated as follows.

HARA utility function is supposed to be the general economical utility function. It has reams of favorable property such as: continuity, monotone increasing and convexity (diminishing marginal utility), which fits the utility function of water pretty well. Furthermore, it is able to represent almost all kinds of functions via merely three parameters, simplifying the fitness of the parameters and enhancing the significance in statistics. Therefore, utilizing HARA to deal with the theoretical analysis of water trading mechanism is a fabulous approach. This study compares the common pool method and pair-wise trading through the HARA utility functions, analyzing the economic efficiency of common pool method in theory.

The main model using HARA utility function is stated in Section 2, with parameters estimated. And the economic advantages of the common pool method are demonstrated in Section 3.

## 2. The Conceptual Generalization of Common Pool Method

*2.1. Express the utilty via HARA utility function*

Water market is a real-time trading taking place in relatively closed drainage basin, participants use water mainly for irrigation. Generally, a participant decides his status (buyer, seller or outsider) by maximizing the utility function $U(w)$, where $w$ is the water the participant has. Assume that the water can't be reserved, so $U(w)$ is determined by two parts: $U_{tr}(w)$ standing for the profit gained by selling the superfluous water and $U_{ag}(w)$ representing the agricultural profit,

$$U(w) = U_{tr}(w) + U_{ag}(w), \qquad (1)$$

Here $U_{tr}(w)$ follows the exponent law and can be written as:

$$U_{tr} = w_{tr} q \cdot e^{\lambda T}, \qquad (2)$$

where $q$ is the water price, $T$ is the growing period of the crops, $\lambda$ is the risk-free interest rate, and $w_{tr}$ is the amount of water used for trading. The participant sells water if $w_{tr} > 0$, while buys water if $w_{tr} < 0$. Here transportation cost is added to the water price and the risk of default is ignored for convenience.

While $U_{ag}(w)$ is some kind of non-negative valued function which satisfies the properties of continuity, monotone increasing, convexity and boundedness. Recommended by the Borch's theorem [28], the so-called HARA (Hyperbolic Absolute Risk Aversion) utility function [29] is supposed to be the general economical utility function. It can be written as:

$$U_{ag} = \frac{1-\gamma}{\gamma} \left(\frac{a \cdot w_{ag}}{1-\gamma} + b\right)^{\gamma} \cdot p_{cr}, \qquad (3)$$
$$= Y_{cr}(w) \cdot p_{cr}$$

where $a$ is the participant's utilization of water resource, the intercept $b$ engenders from other factors besides water, $\gamma$ can be regarded as the type of crops' impact on the utility function, $p_{cr}$ is the price of the corresponding crop, and $Y_{cr}(w)$ is the yield of the crop.

In practice, participants use water mainly for the agricultural production, and make investment when the initial water amount is superfluous or insufficient. Noting the proportionality coefficient as $\alpha$, where $\alpha > 1$ corresponds to buyers and $0 \leq \alpha < 1$ corresponds to sellers. So the utility function can be rewritten as:

$$U(w) = U_{ag} + U_{tr}$$
$$= \frac{1-\gamma}{\gamma} \left(\frac{a \cdot \alpha w}{1-\gamma} + b\right)^{\gamma} \cdot p_{cr} + (1-\alpha)wq \cdot e^{\lambda T}.$$

(4)

Invoking the relationship between the actual diverted water per year (w) and the yield of wheat and utilizing nonlinear fitting in statistics, we are able to acquire the curve of HARA utility function in Figure 1:

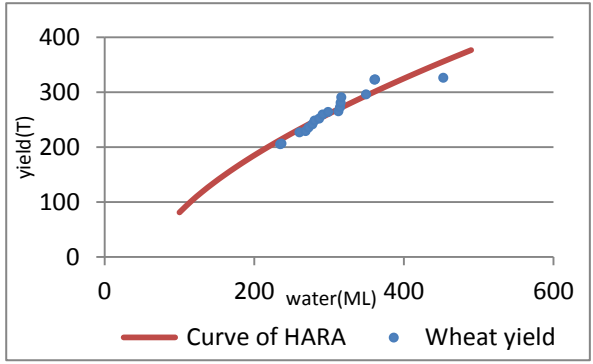

FIGURE 1: The curve of HARA utility function and the actual yield of wheat

The fitting result indicates that the model fits the actual data well, thereby, HARA utility function is applicable to measure the economic benefits of water utility.

*2.2. The optimal mechanism of the common pool method*

A rational participant will maximize his utility function $U(w)$ by

$$\max U(w) = \max_{\alpha \in [0, +\infty]} [\frac{1-\gamma}{\gamma}(\frac{a \cdot \alpha w}{1-\gamma} + b)^\gamma \cdot p_{cr} + (1-\alpha)wq \cdot e^{\lambda T}]$$

(5)

or equally,

$$\max_{w_{tr}, w_{ag}} [w_{tr} q \cdot e^{\lambda T} + \frac{1-\gamma}{\gamma}(\frac{a \cdot w_{ag}}{1-\gamma} + b)^\gamma \cdot p_{cr}]$$
$$S.T. \quad w_{tr} + w_{ag} = w$$

(6)

and it can be solved with the Lagrangian function:

$$L = \frac{1-\gamma}{\gamma}(\frac{a \cdot w_{ag}}{1-\gamma} + b)^\gamma \cdot p_{cr} + w_{tr} q \cdot e^{\lambda T} - m \cdot (w_{tr} + w_{ag} - w)$$

(7)

The corresponding Gradient equation set for this function is as follows:

$$\begin{cases} a(\frac{a \cdot w_{ag}}{1-\gamma} + b)^{\gamma-1} \cdot p_{cr} = m \\ q \cdot e^{\lambda T} = m \\ w_{ag} + w_{tr} = w \end{cases}$$

(8)

with the solution

$$\begin{cases} w_{ag} = [(\frac{q \cdot e^{\lambda T}}{a \cdot p_{cr}})^{\frac{1}{\gamma-1}} - b] \cdot \frac{1-\gamma}{a} \\ w_{tr} = w - [(\frac{q \cdot e^{\lambda T}}{a \cdot p_{cr}})^{\frac{1}{\gamma-1}} - b] \cdot \frac{1-\gamma}{a} \end{cases}$$

(9)

So each participant will maximize his welfare by

$$\max_{w_{tr,i}, w_{ag,i}} w_{tr,i} q \cdot e^{\lambda T} + \frac{1-\gamma}{\gamma}(\frac{a_i \cdot w_{ag,i}}{1-\gamma} + b_i)^\gamma \cdot p_{cr}$$
$$i = 1, 2, \ldots, n$$
$$S.T. \begin{cases} w_{tr,i} + w_{ag,i} = w_i & i = 1, 2, \ldots n \\ \sum_{i=1}^{n} w_{tr,i} = 0 \end{cases}$$

(10)

where $i$ stands for the $i$th participant, and $q$ is the equilibrium price of the market. Furthermore, $\gamma$ can be seen identically for the corps in the same drainage basin are similar. Owing to the same risk-free interest rate in this distinct, $\lambda$ is also identical.

The initial boundary condition is

$$\sum_{i=1}^{n} w_{tr,i} + \sum_{i=1}^{n} w_{ag,i} = \sum_{i=1}^{n} w_i = W, \quad (11)$$

where $W$ is the total water allocation in the water drainage. The unique solution of Equations (10) and (11) is

$$\begin{cases} w_{ag,i} = [(\frac{q \cdot e^{\lambda T}}{a_i \cdot p_{cr}})^{\frac{1}{\gamma-1}} - b_i] \cdot \frac{1-\gamma}{a_i} \\ w_{tr,i} = w_i - [(\frac{q \cdot e^{\lambda T}}{a_i \cdot p_{cr}})^{\frac{1}{\gamma-1}} - b_i] \cdot \frac{1-\gamma}{a_i} \end{cases}$$

$$with\ q = [\frac{\frac{W}{1-\gamma} + \sum_{i=1}^{n}\frac{b_i}{a_i}}{\sum_{i=1}^{n}(\frac{1}{a_i})^{\frac{1}{\gamma-1}}}]^{\gamma-1} \cdot \frac{p_{cr}}{e^{\lambda T}}.$$

(12)

*2.3. An empirical example from Australia*

Now that we have got the relationship between the total water allocation ($W$) and the water price($q$), it is possible to identify if the price of the water is overestimated or underestimated. Suppose $\lambda = 0.06$, which is based on the real risk-free rate in Australia, $P_{cr}$ is wheat price shown in Table 1, which is based on the actual price($) of wheat per ton from ASX (Australian Securities Exchange), and the number of participants $n = 100 \times 15\%$, which is according to the data from Victoian Water Register(http://waterregister.vic.gov.au/). Table 1 indicates the relationship between the water price and the water allocation.

TABLE 1: The relationship between the water price and the water allocation from July 2015 to June 2016 in Murray, Australia

| Month | Water(GL) | Actual median price($/ML) | Wheat price($/T) | Price in the model ($/ML) | Relative residual |
|---|---|---|---|---|---|
| JUL | 37 | 260 | 280 | 262.6 | 1% |
| AUG | 54.3 | 200 | 260 | 210.4 | 5% |
| SEP | 75.4 | 200 | 290 | 206.9 | 3% |
| OCT | 34.2 | 245 | 270 | 261.1 | 7% |
| NOV | 46.3 | 280 | 300 | 258.0 | 8% |
| DEC | 76.6 | 270 | 285 | 202.1 | 25% |
| JAN | 40.4 | 255 | 280 | 253.8 | 0% |
| FEB | 62.5 | 210 | 270 | 207.0 | 1% |
| MAR | 78.9 | 230 | 265 | 185.8 | 19% |
| APR | 64.3 | 225 | 275 | 208.5 | 7% |
| MAY | 51.2 | 245 | 285 | 235.8 | 4% |
| JUN | 59.6 | 185 | 250 | 195.2 | 5% |

Conspicuously, the price in the model accords with actual median price well, suggesting our analysis is logically acceptable. It is effortless to identify that except December and March, most of the time the relative error is less than 10%. The abnormity in December is easy to understand, for it is during the filling stage of wheat, which indicates the demand of water is much higher than uaual. However, our model does not take this factor into consideration, resulting in the relative higher error. The error in May, nonetheless, is difficult to explain. A possible explanation is that other crops' influence is indispensible, hence, the wheat itself has no capacity to determine the water price. For instance, maybe we ought to take cotton into account, which is also a significant crop in Australia.

## 3. The Economic Efficiency of Common Pool Method

The economic efficiency of common pool are analyzed as follows.

Conspicuously, according to the optimal conditions (10), as long as others don't change tactics, each participant's utility is maximized, and they will not change their strategies in as much as it will lower their utilities. Thereby, the solution (12) is a Nash Equilibrium which will guarantee the market's stability[30]. Besides, there exists other advantages as follows.

Proposition I.

The overall welfare of society is improved compared with pair-wise trading.
Proof.

The overall welfare of the society is defined as follows:

$$U_{all} = \sum_{i=1}^{n} U_i. \qquad (13)$$

For common pool, according to (12), the overall welfare of the society is:

$$\sum_{i=1}^{n} \max_{w_{tr,i}, w_{ag,i}} \left( \frac{1-\gamma_i}{\gamma_i} \left( \frac{a_i \cdot w_{ag,i}}{1-\gamma_i} + b_i \right)^{\gamma_i} + w_{tr,i} q \cdot e^{\lambda T} \right)$$

$$S.T. \begin{cases} w_{tr,i} + w_{ag,i} = w_i & i = 1, 2, \ldots n \\ \sum_{i=1}^{n} w_{tr,i} = 0. \end{cases}$$

(14)

For pair-wise trading, two participants trade privately, and their goals are to maximize their own utility function. Supposing two participants $i$ and $j$ make a deal, then their optimal problem can be written as follows:

$$\max_{w_{tr,k}, w_{ag,k}} w_{tr,k} \tilde{q} \cdot e^{\lambda T} + \frac{1-\gamma}{\gamma} \left( \frac{a_k \cdot w_{ag,k}}{1-\gamma} + b_k \right)^{\gamma}$$

$$k = i, j$$

$$S.T. \begin{cases} w_{tr,k} + w_{ag,k} = w_i & k = i, j \\ \sum_{k=i,j} w_{tr,k} = 0, \end{cases}$$

(15)

where $\tilde{q}$ is the optimal price determined by these two participants. The generation mechanism is just the same as (12).

If we use this model to trade in the market, then for the entire market, the overall welfare is:

$$\sum_{i=1}^{n} \max_{w_{tr,i}, w_{ag,i}} \left( \frac{1-\gamma_i}{\gamma_i} \left( \frac{a_i \cdot w_{ag,i}}{1-\gamma_i} + b_i \right)^{\gamma_i} + w_{tr,i} \tilde{q}_i \cdot e^{\lambda t} \right)$$

$$S.T. \begin{cases} w_{tr,i} + w_{ag,i} = w_i & i = 1, 2, \ldots n \\ w_{tr,2k-1} + w_{tr,2k} = 0 & k = 1, 2, \ldots \frac{n}{2} \\ \sum_{i=1}^{n} w_{tr,i} = 0. \end{cases}$$

(16)

Compared with (14), (16) adds a constraint condition:

$$w_{tr,2k-1} + w_{tr,2k} = 0 \qquad k = 1, 2, \ldots \frac{n}{2}. \qquad (17)$$

According to the theory in programming problem in operational research which shows that the optimal solution will not increase if new constraint is added. Therefore, the efficiency under pair-wise trading is inferior to the common pool method. □

Proposition II.

The market reaches the Pareto optimal (i.e. the state of allocation of resources in which it is impossible to make any one individual better off without making at least one individual worse off) which will maximize the overall welfare of the whole market.
Proof.

The strategy of a participant (9) gains Pareto optimal for each participant without considering the market clearing condition (11), for it realizes the maximum of the individual utility.

In order to demonstrate the proposition in the whole market, we should show that any further trading based on solution (12) would not engender more efficiency. Assuming the trading amount is $d$ ($d>0$), then

$$\begin{cases} \Delta U_1 = -dq + \frac{1-\gamma}{\gamma} \left( \frac{a_1 \cdot (w_1 + d)}{1-\gamma} + b_1 \right)^{\gamma} \\ \quad - \frac{1-\gamma}{\gamma} \left( \frac{a_1 \cdot w_1}{1-\gamma} + b_1 \right)^{\gamma} \\ \Delta U_2 = dq + \frac{1-\gamma}{\gamma} \left( \frac{a_2 \cdot (w_2 - d)}{1-\gamma} + b_2 \right)^{\gamma} \\ \quad - \frac{1-\gamma}{\gamma} \left( \frac{a_2 \cdot w_2}{1-\gamma} + b_2 \right)^{\gamma}. \end{cases}$$

(18)

Let

$$f(d) = \Delta U_1 + \Delta U_2, \qquad (19)$$

so $f(0) = 0$, and

$$f'(d) = a_1(\frac{a_1 \cdot (w_1 + d)}{1 - \gamma} + b_1)^{\gamma - 1}$$
$$-a_2(\frac{a_2 \cdot (w_2 - d)}{1 - \gamma} + b_2)^{\gamma - 1} \quad . \quad (20)$$

Invoking the intermediate result of the Lagrange multiplier method:

$$a_i(\frac{a_i \cdot w_{a,i}}{1 - \gamma_i} + b_i)^{\gamma_i - 1} = m \quad i = 1, 2, \ldots n. \quad (21)$$

Therefore, $f'(0) = 0$. Similarly, by the characteristics of HARA utility function, we get $u'' < 0$, thereby $f''(d) < 0$, and we can see that $f(d) < 0 (for\ all\ d > 0)$, which means there exists no $d$ $(d>0)$, such that $f(d) > 0$, which indicates solution (12) is a Pareto optimal. □

Proposition III.

The efficiency of the market is largely increased in comparison with pair-wise trading.
Proof.

In accordance with proposition III, the common pool method can reach the Pareto optimal price at a time, while the pair-wise trading will not. Instead, we introduce a concept in economy called stable matching which a weaker than Pareto optimal to describe the efficiency of pair-wise trading and will illustrate that the stages to find the stable matching is much more compared with common pool method.
Stable matching: A deal is called stable matching [31,32] if there is no such combination: two buyers $\alpha$ and $\beta$ who are trading with two sellers A and B, respectively, although $\beta$ prefers A to B and A prefers $\beta$ to $\alpha$.
It can be proved that in the pair-wise trading market, a stable matching will reach in at most $n^2 - 2n + 2$ stages [33].

Therefore, it's clear that the efficiency of common pool is much higher than the pair-wise trading. □

## 4. Conclusions

A conceptual mathematical model of the common tool water market is constructed in this paper, with the HARA utility function adopted. Theoretical analysis is taken to demonstrate the common tool method is better than the pair-wise trading in many aspects, e.g. the overall welfare of society is improved, the market reaches the Pareto optimal and the efficiency of the market is largely increased. Empirical analysis is done in two steps. Firstly, the parameters are estimated to verify the rationality of using HARA-typed utility function in water market. Furthermore, the bargain value is estimated using the data of Australian water market. It is effortless to identify that most of the time the relative error is less than 10%.

## Conflict of Interest

The authors declare that there is no conflict of interest regarding the publication of this paper.

## References

1. M. Heydari, F. Othman and K. Qaderi, "Developing Optimal Reservoir Operation for Multiple and Multipurpose Reservoirs Using Mathematical Programming," *Mathematical Problems in Engineering*, 2015.
2. H.N. Turral, T. Etchells and H.M.M. Malano, "Water trading at the margin: The evolution of water markets in the Murray‐Darling Basin," *Water Resources Research*, vol. 41, no. 7, 2005.
3. C.W. Howe, D.R. Schurmeier and W.D. Shaw, "Innovative Approaches to Water Allocation: The Potential for Water Markets," *Water Resources Research*, vol. 22 no. 4, pp. 439-445,

1986.

4. H.J. Vaux and R.E. Howitt, "Managing Water Scarcity: An Evaluation of Interregional Transfers," *Water Resources Research*, vol. 20, no. 7, pp. 785-792, 1984.

5. E. Hadjigeorgalis and J. Lillywhite, "The impact of institutional constraints on the Limarí River Valley water market," *Water Resources Research*, vol. 40, no. 5, 2004.

6. M. Bekchanov, A. Bhaduri and C. Ringler, "Potential gains from water rights trading in the Aral Sea Basin," *Agricultural Water Management*, vol. *152*, pp. 41-56, 2015.

7. A.C. Jordan, "Market-Driven Solutions to Economic, Environmental, and Social Issues Related to Water Management in the Western USA," *Water*, vol. 1, pp. 19-31, 2009.

8. D.M. Lumbroso, C. Twigger-Ross and J. Raffensperger, "Stakeholders' Responses to the Use of Innovative Water Trading Systems in East Anglia, England," *Water Resources Research*, vol. 28, no. 9, pp. 2677-2694, 2014.

9. J.F. Raffensperger and T.A. Cochrane, "A Smart Market for Impervious Cover", *Water Resources Management*, vol. 24, no. 12, pp. 3065-3083, 2010.

10. J.J. Murphy, A. Dinar and R.E. Howitt, "The Design of 'Smart'' Water Market Institutions Using Laboratory Experiments," *Environmental & Resource Economics*, vol. *17*, no. 4, pp. 375-394 2000.

11. P. Schreinemachers and T. Berger, "An agent-based simulation model of human–environment interactions in agricultural systems," *Environmental Modelling& Software*, vol. 26, no. 7, pp. 845-859, 2011.

12. B. Fisher, K. Kulindwa and I. Mwanyoka, "Common pool resource management and PES: Lessons and constraints for water PES in Tanzania," *Ecological Economics*, vol. 69, no. 6, pp. 1253-1261, 2010.

13. S. Wheeler, A. Loch and A. Zuo, "Reviewing the adoption and impact of water markets in the Murray–Darling Basin, Australia," Journal of Hydrology, vol. *518*, pp. 28-41, 2014.

14. M. ZamanA, H.M. Malano and B. Davidson, "An integrated water trading-allocation model, applied to a water market in Australia," *Agricultural Water Management*, vol. *96*, no. 1, pp. 149-159, 2009.

15. A. Zuo, C. Nauges and S.A. Wheeler, "Farmers' exposure to risk and their temporary water trading," *European Review of Agricultural Economics*, vol. 42, no. 1, pp. 1-24, 2015.

16. T. Etchells, H.M. Malano and T.A. McMahon, "Overcoming third party effects from water trading in the Murray–Darling Basin," *Water Policy*, vol. *8*, no.1, pp. 69-80, 2006.

17. D. Connell and R.Q. Grafton, "Water reform in the Murray‐Darling Basin," *Water Resources Research*, vol. 47, no. 12, 2011.

18. T. Erfani, O. Binions and J.J. Harou, "Simulating water markets with transaction costs," *Water Resources Research*, vol. 50, no. 6, pp. 4726-4745, 2014.

19. S.M.W. Reddy, R.I. Mcdonald and A.S. Maas, "Industrialized Watersheds Have Elevated Water Risk and Limited Opportunities to Mitigate Risk through Water Trading," *Water Resources & Industry*, vol. *1*, pp. 27–45, 2015.

20. J. Yong, "China's water scarcity," *Journal of Environmental Management*, vol. 90, no. 11, pp. 3185-3196, 2009.

21. X. Zeng, Y. Li, G. Huang and L. Yu, "Inexact Mathematical Modeling for the Identification of Water Trading Policy under Uncertainty," *Water*, vol. *6*, pp. 229-252, 2014.

22. S. Moore, "The Development of Water Markets in China: Progress, Peril, and Prospects," *Water Polic*, 2014.

23. S. Zhong, L. Shen and J. Sha, "Assessing the Water Parallel Pricing System against Drought in China: A Study Based on a CGE Model with Multi-Provincial Irrigation Water," *Water*, vol. *7*, pp. 3431-3465, 2015.

24. Z. Alec, B. Robert, A.W. Sarah, H. Edwyna and

B. Henning, "Understanding Irrigator Bidding Behavior in Australian Water Markets in Response to Uncertainty", *Water*, vol. *6*, no. 11, pp. 3457-3477, 2014 .

25. S.B. Miguel, "Water Institutions and Management in Cape Verde," *Water*, vol. 7, pp. 2641-2655, 2015.

26. S.O. Lopes, F., A.C.C. Fernando, P. and M.S. Rui, "Optimal Control Applied to an Irrigation Planning Problem," *Mathematical Problems in Engineering*, 2016.

27. H. Chang and X. Rong, "Legendre Transform-Dual Solution for a Class of Investment and Consumption Problems with HARA Utility," *Mathematical Problems in Engineering*, 2014.

28. C. Hao, C. Kai, L. Ji-mei and M. Simone, "Portfolio Selection with Liability and Affine Interest Rate in the HARA Utility Framework," *Abstract and Applied Analysis*, *2014*.

29. L. Markus, "Calculation of Price Equilibria for Utility Functions of the HARA Class," *Astin Bulletin*, vol. 16, no. S1, pp. 91-97, 1986.

30. Z. Wu, C. Dang and H. Karimi, "A Mixed 0-1 Linear Programming Approach to the Computation of All Pure-Strategy Nash Equilibrium of a Finite n-Person Game in Normal Form," *Mathematical Problems in Engineering*, 2014.

31. Á. and F.M. David, "Stable marriage and roommates problems with restricted edges: Complexity and approximability," *Discrete Optimization*, 2016.

32. D. Gale and L.S. Shapley, "College Admissions and the Stability of Marriage," *American Mathematical Monthly*, vol. *69*, pp. 9-15, 1962.

33. E. Avlos, M. Dimitrios, M. Ioannis and M. Panayiotis, "Finding a minimum-regret many -to-many Stable Matching," *Optimization*, vol. 62, no. 8, pp. 1007-1018, 2013 .